\documentclass[10pt,twocolumn]{article} 

\usepackage{latex8}
\usepackage[utf8]{inputenc}
\usepackage{amssymb}
\usepackage{url}

\usepackage[compact]{titlesec}
\titlespacing{\section}{0pt}{2ex}{1.5ex}
\titlespacing{\subsection}{0pt}{1ex}{1ex}
\titlespacing{\subsubsection}{0pt}{0.5ex}{0ex}

\begin{document}

\title{Supporting Parallelism in Server-based Multiprocessor Systems}

\author{Luís Nogueira, Luís Miguel Pinho\\
CISTER Research Centre\\
School of Engineering (ISEP), Polytechnic Institute of Porto (IPP), Portugal\\
\{lmn,lmp\}@isep.ipp.pt}

\maketitle

\begin{abstract}
Developing an efficient server-based real-time scheduling solution that supports dynamic task-level parallelism is now relevant to even the desktop and embedded domains and no longer only to the high performance computing market niche. This paper proposes a novel approach that combines the constant-bandwidth server abstraction with a work-stealing load balancing scheme which, while ensuring isolation among tasks, enables a task to be executed on more than one processor at a given time instant. 
\end{abstract}

\section{Introduction}

The constant-bandwidth server abstraction has proved very useful in designing, implementing, and reasoning about single core open real-time systems where tasks can dynamically enter or leave the system at any time \cite{luisJSA10}. Each task is assigned a fraction of the computational resources and it is handled by an abstract entity called server to achieve the goals of temporal isolation and real-time execution \cite{abeni98}.

However, modern open real-time systems increasingly generate heavy workloads and it is rapidly becoming unreasonable to expect to implement them as single core systems. In fact, a general shift from unicore to multicore processors can be seen both in the general purpose and embedded domains as an energy-efficient way to boost applications' performance. Therefore, there have been significant efforts to extend reservation-based real-time scheduling theory to make it applicable to multiprocessor systems as well \cite{baruah02,pellizzoni08,faggioli10,kato10}. 

Nevertheless, all these works consider task models where tasks use at most a single core at each time instant. This restriction is natural for uniprocessor scheduling since only one processor is available at any time, even if we deal with parallel algorithms. However, the need for parallel processing - simultaneous use of several processors for an individual task - is steadily increasing, even in the desktop and embedded domains and no longer only on the comparably small high performance computing market niche. Therefore, for fully utilising the parallel abilities of multicore platforms, we should be able to support tasks that may be executed on different cores at the same time instant.

There are many computations that can be relatively easily parallelised by using frameworks such as Cilk \cite{frigo98}, Intel's Parallel Building Blocks \cite{intelpbb}, Java Fork-join Framework \cite{lea00}, Microsoft's Task Parallel Library \cite{microsofttpl}, or StackThreads/MP \cite{taura99}. These frameworks encourage application developers to create many more parallel jobs (hereafter called \emph{pjobs}) than there are available CPUs. The division of work among pjobs is often imperfect, and the system must provide an efficient run-time that can efficiently map ready pjobs to processors, thus dynamically balancing the workload. 

One of the simplest, yet best-performing, dynamic load-balancing algorithms for shared-memory architectures is work-stealing \cite{blumofe99}. Blumofe and Leiserson have theoretically proven that the work-stealing algorithm is optimal for scheduling fully-strict computations \cite{blumofe99}. Under this assumption, an application running on $P$ processors achieves $P$-fold speedup in its parallel part, using at most $P$ times more space than when running on one CPU. These results are also supported by experiments \cite{saha07}.

This paper discusses the general guidelines of a novel scheduling approach for parallel runtimes that will coexist with a wide range of other complex independently developed applications, without any previous knowledge about their real execution requirements, number of pjobs, and when those pjobs will be generated. Schedulers in these type of open systems are therefore required to maintain a certain (quantifiable) level of service for each application, with the exact guarantee depending upon the CPU reservation's parameters. The proposed approach combines a work-stealing policy with multiprocessor constant-bandwidth servers which, while ensuring isolation among tasks, allows a task to be executed in more than one processor at a given time. To the best of our knowledge, no research has ever focused on this subject.

\section{System model}

We consider the scheduling of sporadic independent tasks on $m$ identical processors ${p_1, p_2, \ldots, p_m}$ using global EDF. With global EDF, each task ready to execute is placed in a system-wide queue, ordered by nondecreasing absolute deadline, from which the first $m$ tasks are extracted to execute on the available processors. 

A pool of worker threads is established. Assume that there will be as many worker threads as there are CPUs on a system. A special purpose accounting discipline is used to manage tasks and execute them via the worker threads.

Each task $\tau_i$ can generate a virtually infinite sequence of jobs. The arrival time $a_{i,j}$ of the $j^{th}$ job of a task $\tau_i$ is only revealed at runtime and the exact execution requirements $e_{i,j}$ can only be determined by actually executing the job to completion until time $f_{i,j}$. All jobs generated by a task $\tau_i$ are dedicated to a p-CSS server $S_i$, an extension for the multicore case of the Capacity Sharing and Stealing scheduler \cite{luisJSA10}. Each server $S_i$ is characterised by a pair $(Q_i,T_i)$, where $Q_i$ is the server's maximum reserved capacity and $T_i$ its period. The ratio $U_i = \frac{Q_i}{T_i}$ denotes the fraction of the capacity of one processor that is assigned to the server. 

At each instant, the following values are associated with a server $S_i$: its currently assigned deadline $d^i_k$, its remaining execution capacity $0 \le c^i_k \le Q_i$, the amount of residual capacity $r^i_k \le c^i_k$ that can be reclaimed by other servers, and its currently assigned replenishment time $h^i_k = d^i_k$. If at time $t$, $S_i$ finishes the execution of its currently served job without exhausting its reserved execution capacity $c^i_k$ and it has no pending work, the remaining amount $c^i_k > 0$ sets the server's residual capacity $r^i_k = c^i_k$ that can be reclaimed by other servers ($c^i_k$ is subsequently set to zero). By pending work we refer to the case when there exists at least a served job such that its release time is $s_{i,j} \le t < f_{i,j}$. 

During the course of its execution a job can spawn, at any time, a set of parallel jobs $\{pjob_{i,1},pjob_{i,2},\ldots,pjob_{i,n}\}$, sequential pieces of work that can be executed on different processors at the same time instant using the available execution capacity of their corresponding task. For now, our work is focused on systems where all pjobs are fully independent, \emph{i.e.}, except for the $m$-cores there are no other shared resources, no critical sections, nor precedence constraints.

Contrary to regular jobs of a task, pjobs are not pushed to the global EDF queue but instead maintained in a worker's local work-stealing double-ended queue (deque) to reduce contention on the global queue. Any pjob in the work-stealing queue can be shared with any other worker thread. A worker thread first looks into its local queue. If there is no pjob to pick, then it searches the global EDF queue. Still, if there is no eligible job\footnote{An eligible job $j_{i,j}$ is one in which its dedicated server $S_i$ is able to execute $j_{i,j}$ by either consuming its own reserved capacity $c^i_j > 0$, reclaiming any available capacity $r^i_k$ with deadline $d^i_j \le d^i_k, r^i_k > 0$, or stealing a non-isolated capacity $c^s_k$ with deadline $d^s_k \leq d^i_j, c^s_k > 0$} in the global EDF queue, the worker will steal the earliest deadline eligible pjob from the top of other busy worker's deque. For a busy worker, pjobs are pushed and popped from the bottom of the deque and these operations are synchronisation-free.

\section{Multicore Capacity Sharing and Stealing}

In this paper, we consider a periodic task model in which jobs may spawn a set of parallel jobs, independent sequential pieces of code that may have different execution costs but a common period. Multithreaded jobs such as this arise naturally in many settings. For example, in multimedia applications, multiple threads may be useful for performing different functions on common data (e.g., a frame of an MPEG video) at the same rate. Our goal is to find an efficient scheduling framework for these parallel runtimes while ensuring temporal isolation among applications and guaranteeing a certain degree of service to each individual application in open real-time systems. 

Since our management of reserved capacities is based on our previous work for uniprocessor systems \cite{luisJSA10}, we will start by describing the capacity sharing and stealing approach of CSS. CSS extends CBS \cite{abeni98} with a powerful strategy that supports the coexistence of guaranteed (\emph{isolated}) and non-guaranteed (\emph{non-isolated}) bandwidth servers to efficiently handle soft-tasks' overloads by making additional capacity available from two sources: (i) reclaiming unused reserved capacity when jobs complete in less than their budgeted execution time and (ii) stealing reserved capacity from inactive non-isolated servers used to schedule best-effort jobs.

Whenever a job is being executed, the consumed execution capacity must be decreased by the same amount. By dynamically managing a pointer to the server from which the capacity is going to be decreased, the proposed dynamic accounting mechanism of CSS eliminates the need of extra queues or additional server states, reducing its overhead. The server from which the accounting is going to be performed is dynamically determined at the time instant when a capacity is needed. CSS uses the following rules to manage reserved capacities:

\begin{itemize}
\item\textbf{Rule A (residual capacity release):} Whenever a server $S_j$ completes its $k^{th}$ job of its associated task $\tau_j$ and it has no pending work, its remaining reserved capacity $c^j_k > 0$ is released as residual capacity $r^j_k = c^j_k$ and $c^j_k$ is set to zero. The released residual capacity $r^j_k$ can immediately be reclaimed by eligible active servers until the currently assigned $S_j$'s deadline $d^j_k$. $S_j$ is kept active with its current deadline until its residual capacity $r^j_k$ is exhausted by other servers.
\item\textbf{Rule B (residual capacity reclaim):} The next active server $S_i$ scheduled for execution points to the earliest deadline server $S_{edf}$ from the set of eligible active servers $A_r$ for capacity reclaiming. $S_i$ consumes the pointed residual capacity $r^{edf}_k$, running with the deadline $d^r_k$ of the pointed server $S_{edf}$. Whenever $r^{edf}_k$ is exhausted and there is pending work, $S_i$ disconnects from $S_{edf}$ and selects the next available server $S_{edf}'$ (if any).
\item\textbf{Rule C (dedicated capacity consumption):} If all eligible residual capacities are exhausted and the current $k^{th}$ job of server $S_i$ is not yet completed, $S_i$ consumes its own reserved capacity $c^i_k$ either until the job's completion or $c^i_k$'s exhaustion (whatever comes first). If $c^i_k$ is exhausted and there is still pending work to do, $S_i$ is kept active with its current deadline $d^i_k$.
\item\textbf{Rule D (inactive non-isolated capacity steal):} A server $S_i$ with pending work and no available execution capacity ($c^i_k = 0$) connects to the earliest deadline server $S_{edf}$ from the set of eligible inactive non-isolated server $I_s$. $S_i$ steals the pointed inactive capacity $c^{edf}_k$, running with its current deadline $d^i_k$. Whenever $c^{edf}_k$ is exhausted and the job has not yet been completed, the next non-isolated capacity $c^{edf'}_k$ is used (if any). 
\end{itemize}

We are currently investigating how these rules can be extended for multicore platforms. Due to well-known multiprocessor scheduling anomalies \cite{andersson02}, the extension is not trivial and adopting the same rules as the uniprocessor case would lead to deadline violations in spite of the fact that the considered task set is schedulable by using a global EDF scheduler. A possible approach for reclaiming residual capacities has been proposed in M-CASH \cite{pellizzoni08} (residual capacities are equally distributed across all processors, including idle ones) but no work is known for handling capacity stealing in the multicore case. 

\section{Work stealing in the presence of jobs' priorities}

Work-stealing schedulers are increasing in popularity as scheduling algorithms for dynamic task parallelism. A work-stealing scheduler employs a fixed number of threads called workers. Each of those workers has a local deque to store tasks. Whenever a worker has no local tasks to execute it will try to steal a task from the top of other busy worker's deque. Thus, it must choose which processor will be stolen and which task will be taken. These choices lead to variations of the work-stealing algorithm and are the main issue of this section.

Blumofe and Leiserson \cite{blumofe99} demonstrate that a random choice of the stolen processor is fair. Furthermore, random choices present the advantage that the choice of the target does not require more information than the total number of processors in the executive platform. Then, the thread steals a task from the back of the run-queue of the randomly chosen thread. The reasons for accessing the run queues at different ends are several \cite{frigo98}: (i) it reduces contention by having stealing threads operate on the opposite end of the queue than the thread they are stealing from; (ii) it works better for parallelised divide-and-conquer algorithms which typically generate large chunks of work early, so the older stolen task is likely to further provide more work to the stealing thread; and (iii) stealing a task also migrates its future workload, which helps to increase locality. All queue manipulations run in constant-time (O(1)), independently of the number of tasks in the queues.


However, the algorithm does not take tasks' deadlines into account when stealing a task from another worker. Also, a task must voluntarily yield or block before another task can be scheduled on the same CPU or otherwise it will run to completion. Thus, the need to support tasks' deadlines and CPU reservations fundamentally distinguishes the problem at hand in this paper from other work-stealing choices previously proposed in the literature.

Our proposal is to apply work-stealing to enable multithreaded jobs to be executed on more than one processor. Recall that a job $j_{i,k}$ assigned to a particular worker thread by a global EDF policy, can spawn a set of pjobs at any time during the course of its execution. Pjobs are not pushed to the global EDF queue but instead pushed to the bottom of the worker's local deque. Pjobs are dedicated to same server $S_i$ of job $j_{i,k}$, ensuring isolation among tasks. 

Then, while there are pending pjobs on the local deque, the worker should purposefully select the bottommost pjob, which is the pjob with the highest probability of still being in the cache. Hence, there are performance improvements in processing local queues in a LIFO order. Note that in this case the job is sequentially executed as if there was no support for parallel execution of pjobs. On the other hand, work stealing allows an idle worker thread to perform some of the pjobs in other overloaded processor's queue. Thus, whenever a worker thread has no pending pjobs in its local deque and the first pjob on the global EDF queue has a greater deadline than at least one of the eligible pjobs at the top of the other workers' deques, the worker thread should steal the earliest deadline eligible pjob from the topmost pjobs on the other workers' deques. 

We believe that this deadline-based work-stealing policy will positively increase the speedup of parallel applications without jeopardising the schedulability of the other sequential jobs scheduled by global EDF. We are currently investigating such claim. 

\section{Conclusions and future work}

This paper discussed the increased need to support dynamic task-level parallelism in open real-time systems and proposed the general guidelines of a novel scheduling approach that combines a work-stealing load balancing policy with a multicore reservation-based approach.  

Our current efforts are focused on a theoretical validation of the proposed approach. It is our belief that the ideas discussed here will improve the execution efficiency of parallel tasks while continuing to achieve isolation among tasks whose resource demands are only know at runtime. We plan to evaluate the efficiency of the approach in real-world scenarios by implementing it on top of SCHED\_DEADLINE \cite{faggioli09}, a patch made for the Linux kernel which implements an EDF scheduler with a CPU reservation mechanism based on CBS.

\section*{Acknowledgements}
This work was supported by FCT through the CooperatES (PTDC/EIA/71624/2006) and SENODs projects (CMU-PT/SIA/0045/2009), and by the European Commission through the ARTIST2 NoE (IST-2001-34820).

\bibliographystyle{latex8}
\bibliography{references.bib}

\end{document}